\documentclass [12pt]{article}

\begin{document}
\title{\bf Electromagnetic Influence on Gravitational Mass -- Theory, 
Experiments, and Mechanism of the Solar Corona Heating}
\author{Miroslav S\'{u}ken\'{\i}k and Jozef \v{S}ima \\[1ex]
  Slovak Technical University, FCHPT, Radlinsk\'{e}ho 9, \\
  812 37 Bratislava, Slovakia
  \\
  e-mail:  sukenik@minv.sk, sima@chtf.stuba.sk} \date{}
\maketitle
\begin{abstract}
\label{sec:abstractne}
Based on the model of Expansive Nondecelerative Universe, 
the aim of the present contribution is to contribute to the theoretical 
rationalization of some experiments, in particular of those performed by 
Podkletnov and de Aquino and devoted to the gravitational mass cessation, to 
offer a mechanistic explanation of the Solar corona heating, and to propose 
an experiment to verify the explanation of the heating in the Earth 
conditions. 


\end{abstract}

\section*{Introduction}

Among the problems of current physics, issues concerning gravity are
still ranked in ``top ten''. Theoretical approaches focused on
elaboration of a theory of quantum gravitation unifying all four (or
more?) fundamental forces are in progress, successful completion of
the theory seems to be, however, far beyond the horizon.
Simultaneously with theoretical attempts (for a general survey, see
[1]), experiments aimed at influencing the gravitational mass by other
forces have been developed and tentatively explained [2-8]. The
gravity has penetrated also into chemistry as manifested by
space-craft experiments (preparation of compounds, alloys or mixtures
that, due to the gravitational force acting at the Earth, cannot be
prepared in the Earth conditions) as well as by a newly elaborated
rationalization of some quantum chemical effects [9-10].

A carefull inspection of the contributions devoted to experiments
involving gravitational effects [2-7] may lead the reader to a
conclusion on the necessity to introduce a theoretical approach
different of those applied in the original papers. It concerns,
namely, an approach based on the general theory of relativity and
benefiting from the ability to localize and quantify the gravitational
energy. The Expansive Nondecelerative Universe (ENU) model [8,11-13]
satisfies the above requirements. The purpose of the present
contribution is to contribute to the rationalization of some
experiments, in particular of those performed by Podkletnov [2-4] and
de Aquino [5-7], to offer a mechanistic explanation of the Solar
corona heating, and to propose an experiment enabling to verify the
explanation in the Earth conditions.

\section*{Podkletnov's phenomenon}

In the papers by Podkletnov [2-4] a detection of an anomalous force 
originating at a high-voltage discharge directed from a superconductive 
emitter to a target electrode is described. The effect has been attributed 
to forces of gravitational nature. At the Podkletnov's experiments, the 
following conditions were applied: the discharge time was 10$^{ - 5}$ to 
10$^{ - 4}$ s, the current peak value at the discharge was of the order 
10$^{4}$ A, the voltage varied from 500 kV to 2 MV, the interelectrode 
distance ranged from 15 cm to 2 m, the total charge of the emitter was about 
0.1 C, and the discharge energy approached $10^5$ J. The formed 
gravitational impulse propagated as a coherent beam in the same direction as 
the discharge. It penetrates through different media, apparently without any 
energy loss and caused a deflection of a 10 - 50 g pendulum. The energy of 
the deflection was independent on the pendulum material and mass, and 
depended on the voltage only. 

One of the possible explanations of the Podkletnov's phenomenon was offered 
in [8]. At the Podkletnov's experiments, a mean intensity of the 
electrostatic field was
\begin{equation}
  \label{eq:1a}
 E \cong 2.5\times 10^6 \mbox{ V m$^{ - 1} $}
\end{equation}
which corresponds to a mean energy density
\begin{equation}
  \label{eq:2a}
 \varepsilon _e \cong 27 \mbox{ J m$^{ - 3 }$} 
\end{equation}
As shown in [8], energy density of gravitational field at the Earth surface 
is
\begin{equation}
  \label{eq:3a}
 \left| {\varepsilon _g } \right| = \frac{R  c^4}{8  \pi  G} = 
\frac{3m_{(Earth)} c^2}{4\pi  a  r_{(Earth)}^2 } = 24.29 \mbox{ J m$^{ - 3 }$} 
\end{equation}
where $R$ is the scalar curvature, $a$ is the gauge factor, $m_{(Earth)} $ 
and $r_{(Earth)} $ are the Earth mass and radius, respectively. Based on~(\ref{eq:2a}) 
and~(\ref{eq:3a}) it follows
\begin{equation}
\label{eq1}
\varepsilon _e \cong \varepsilon _g 
\end{equation}

The Earth gravitational field interferes with the electrostatic field 
created in the Podkletnov's equipment which leads to a local cessation of 
the gravitational field. The corresponding deviation from the vertical 
direction of the gravity causes the observed pendulum deflection. In [8] it 
was proved that the deflection in vertical direction, $h_v $ is described by
\begin{equation}
\label{eq2}
h_v \cong \frac{Q}{g}
\end{equation}
where $g$ is the Earth surface gravitational acceleration (9.80665 m s$^{ - 
2})$ and $Q$ is the total charge of the emitter.

\section*{Solar corona}

The temperature profile of the Sun and its surroundings belongs still to the 
enigmas of physics. At the core of the Sun, the temperature reaches about 15 
million K which is understandable since energy is created in the Sun. 
Passing from the core to the surface the temperature decreases and at the 
Sun's surface it is about 5,500 K. Next part (outside of the chromosphere) 
is corona. There is no source of energy, its temperature, however, is 
extremely high (about 1 million K [14]). The question is, what mechanism is 
applied at the corona heating, preserving thus its temperature? This part 
offers a possible explanation. 

The mean magnetic field intensity at the Sun's surface is [15]
\begin{equation}
  \label{eq:6a}
 \bar {H} \cong 10^2 \mbox{ A m$^{ - 1 }$} 
\end{equation}
and during some processes the intensity may increase in sunspots up to
\begin{equation}
  \label{eq:7a}
  H_{\max } \cong 10^5 \mbox{ A m$^{ - 1 }$}
\end{equation}
A magnetic field intensity with a density identical to the gravitational 
energy density on the Sun's surface is the critical density $H_{crit} $ 
reaching
\begin{equation}
  \label{eq:8a}
 H_{crit} \cong 4\times 10^4 \mbox{ A m$^{ - 1 } $}
\end{equation}
which corresponds to the critical magnetic induction
\begin{equation}
  \label{eq:9a}
 B_{crit} \cong 5\times 10^{ - 2}\mbox{ T} 
\end{equation}

Interference of the gravitational and magnetic fields can lead to 
origination of a gravitational impulse that can increase the kinetic energy 
of the particles forming the Solar corona, i.e. its temperature. The 
mechanism of the corona "heating" may be, therefore, similar to that of the 
Podkletnov's phenomena. The difference is that instead of pendulum and 
electrostatic discharge, rather plasma particles and magnetic impulses are 
in action, i.e. 
\begin{equation}
\label{eq3}
\frac{3m_{(Sun)} c^2}{4\pi  a  r_{(Sun)}^2 } = \frac{B_{crit}^2 }{2\mu _o 
}
\end{equation}

At the Podkletnov's experiments, along with a high voltage, also a
discharge must be created. Similarly, the Sun's magnetic field with
$B_{crit} $ must influence on the particles (electrons) and induce
their precession with the frequency
\begin{equation}
\label{eq4}
\omega _e = \frac{B_{crit} e}{m_e }
\end{equation}
where $m_e $ is the electron mass. In order no avoid damping, the mentioned 
precession must equal to the plasma frequency $\omega _{pl} $ [16] in a 
given region
\begin{equation}
\label{eq5}
\omega _e = \omega _{pl} = \left( {\frac{e^2n_{e(Sun)} }{\varepsilon _o m_e 
}} \right)^{1 / 2}
\end{equation}
where $n_{e(Sun)} $ denotes a concentration of the electrons with a given 
precession frequency in a space. 

In order the Podkletnov's phenomenon may occur, a region with the
concentration of the plasma electrons of $n_{e(Sun)} $ must be just
near the Sun surface where at $B_{crit} $ given by~(\ref{eq:9a}) the electrons
will have identical precession and plasma frequencies. Based on
(\ref{eq3})--(\ref{eq5}) it follows
\begin{equation}
  \label{eq:13a}
 n_{e(Sun)} = \frac{3m_{(Sun)} }{2\pi  m_e a  r_{(Sun)}^2 } \cong 
10^{16} \mbox{ m$^{ - 3}$} 
\end{equation}

Experimental observation of the Sun and its environment has brought an
evidence that such a region (the boundary of chromosphere and corona)
really exists. Thus, just in this region, conditions for the corona
heating by the Podkletnov's phenomenon are satisfied (as a matter of
fact, there is no other place where they might exist).

\section*{A model experiment}

The above conclusions on the conditions required to heat the Solar corona 
may be experimentally verified in a laboratory. Here, such an experiment is 
suggested. The gravitational energy density at the Earth surface (24.29 J 
m$^{ - 3}$ [8]) equals to magnetic field density when
\begin{equation}
  \label{eq:14a}
H_{crit(Earth)} = 7.000\times 10^3 \mbox{ A m$^{ - 1 }$}  
\end{equation}
or
\begin{equation}
  \label{eq:15a}
B_{crit(Earth)} = 8.796\times 10^{ - 3} \mbox{ T}   
\end{equation}
Applying (13) to the conditions at the Earth surface leads to the value of 
concentration of electrons in the plasma with identical the plasma and 
electron precession frequencies 
\begin{equation}
  \label{eq:16a}
 n_{e(Earth)} \cong 5.5\times 10^{14} \mbox{ m$^{ - 3 }$} 
\end{equation}

Such a concentration of free electrons is present in a sample of
hydrogen containing $3.9\times 10^{21}$ hydrogen atoms in a cubic
meter when being at the temperature of 3000 K. This sample is exposed
to a pulsing magnetic field with $B_{crit} $ given by (15). In the
direction of the magnetic poles, a second sample of hydrogen (room
temperature, concentration about $10^{20}$ in a cubic meter) will be
placed next to the first sample. After some time of magnetic field
pulsing, the temperature of the second sample should increase. The
temperature of the control samples placed in other directions should
be preserved.

\section*{Rationalization of experiments performed by de Aquino}

Stemming from the special relativity theory and classical physics, de Aquino 
[17] discussed relations between the inertial and gravitational masses. 
Using a very detailed and precise mathematical treatment he manifested that 
in the absence of electromagnetic field, the masses are identical. In the 
presence of such a field, the gravitational mass may be reduced or even 
nullified. 

In his experiment [5], de Aquino worked with dics-shaped organic luminescent 
material. According to quantum statistical mechanics, an average number of 
photons forming the photon gas inside a luminescent material is given as
\begin{equation}
\label{eq6}
\bar {N}_{(h\nu )} = \left( {\exp (\lambda _e / \lambda _{(h\nu )} ) - 1} 
\right)^{ - 1}
\end{equation}
where $\lambda _e $ and $\lambda _{(h\nu )} $ are the Compton wavelength of 
the electron, 
\begin{equation}
  \label{eq:18a}
 \lambda _e = 2.42\times 10^{ - 12} \mbox{ m}  
\end{equation}
qand a mean wavelength of a photon emitted or absorbed by a luminescent 
material, respectively. 

In the experiment, a material emitting blue light with 
\begin{equation}
  \label{eq:19a}
 \lambda _{(h\nu )} = 461 \mbox{ nm}  
\end{equation}
was used. The radiation output of the material is expressed as
\begin{equation}
\label{eq7}
P_{(h\nu )} = \bar {N}_{(h\nu )} h\nu ^2 = \frac{hc^2}{\lambda _{(h\nu )}^2 
\left( {\exp ((\lambda _e / \lambda _{(h\nu )} ) - 1} \right)}
\end{equation}
Based on~(\ref{eq:18a})--(\ref{eq7}) it can be calculated that for
the radiation of 461 nm wavelength
\begin{equation}
  \label{eq:21a}
 P_{(h\nu )} \cong 56 \mbox{ W}  
\end{equation}
The ENU model provides a general relation [18] for the gravitational output 
of a body with the mass $m$ 
\begin{equation}
\label{eq8}
\left| {P_g } \right| = \frac{d}{dt}\int {\frac{R  c^4}{8\pi  G}dV = 
\frac{m  c^2}{t_c }} 
\end{equation}
where the present cosmological time calculated by the ENU model is [12]
\begin{equation}
  \label{eq:23a}
 t_c = 4.296\times 10^{17} \mbox{ s}  
\end{equation}

A general condition for the interference of the electromagnetic and 
gravitational fields may be expressed as 
\begin{equation}
\label{eq9}
P_{(h\nu )} = \left| {P_g } \right|
\end{equation}
Stemming from (\ref{eq7})--(\ref{eq9}), the ENU provides a total mass of the 
electroluminescent disc 
\begin{equation}
  \label{eq:25a}
 m_{(disc,calc)} \cong 267 \mbox{ kg}  
\end{equation}
In the de Aquino's experiment [5] the discs of a mass
\begin{equation}
  \label{eq:26a}
 m_{(disc,\exp )} \cong 264 \mbox{ kg}  
\end{equation}
were used. Thus, the ENU based calculated and experimental masses
coincide.

If both approaches are consistent, equation (\ref{eq8}) can be exploited to 
calculate a more precise value of the cosmological time and such a 
calculation leads to
\begin{equation}
  \label{eq:27a}
 t_c = 4.169\times 10^{17} ]\mbox{ s} = 1.32\times 10^{10} \mbox{ years}  
\end{equation}

At the same time, the above rationalization leads to a conclusion on a
time dependence of the vacuum permittivity $\varepsilon _o $ and
permeability $\mu _o $.

In the Podkletnov's experiments, the electromagnetic field interferes
with the Earth gravitational field. In case of the experiments carried
out by de Aquino, the electromagnetic field interferes with the
gravitational field of a body.

Relation (\ref{eq9}) is applicable also for the experiment [5] with radiation of 
extra-low frequency (ELF) acting on a ferromagnetic material. It this case 
it must hold
\begin{equation}
\label{eq10}
\eta R_a I^2 = \frac{m_{Fe} c^2}{t_c }
\end{equation}
where $R_a $ is the antenna impedance (116 m$\Omega )$, $I$ is the current 
(130 A), $m_{Fe} $ is the mass of iron powder (29 kg). In such a case the 
absorption coefficient is
\begin{equation}
\label{eq11}
\eta = 3.17\times 10^{ - 3}
\end{equation}
For the iron powder used at the experiment, conductivity gradient $\sigma = 
10$ S m$^{ - 1}$, $\mu = 75\mu _o $ and for the frequency 60 Hz, the skin 
layer thickness is
\begin{equation}
  \label{eq:30a}
 \delta = \left( {\frac{2}{\omega   \sigma  \mu }} \right)^{0.5} \cong 
2.43 \mbox{ m} 
\end{equation}
The thickness of the iron powder then must be
\begin{equation}
  \label{eq:31a}
 d = \eta   \delta \cong 8 \mbox{ mm}  
\end{equation}
which is in an excellent agreement with the experimental value.

The above mentioned ideas may be applied also in the field of 
thermodynamics. In the centre of stars (e.g. the Sun) the gravitational and 
radiation energies must be balanced, i.e.
\begin{equation}
\label{eq12}
\left| {\frac{3m_{(Sun)} c^2}{4\pi  a   r_{g(Sun)}^2 }} \right| = 
\frac{4\sigma _{SB} T_{(Sun)}^4 }{c}
\end{equation}
where $r_{g(Sun)} $ is the gravitational radius of the Sun ($ \cong 3\times 
10^3$ m), $\sigma _{SB} $ is the Stefan-Boltzmann constant ($5.67\times 10^{ 
- 8}$ kg s$^{ - 3}$ K$^{ - 4})$, $T_{(Sun)} $ is the temperature in the 
centre of the Sun. Relation (\ref{eq12}) leads to 
\begin{equation}
  \label{eq:33a}
 T_{(Sun)} \cong 1.5\times 10^7 \mbox{ K} 
\end{equation}
which is identical (within the degree of accuracy of its estimation) to that 
published in various literature sources [e.g. 19]. 

In one of the latest de Aquino's contributions [20], changes in the
gravitational mass of a body due to its interaction with
electromagnetic radiation of very low frequency are described and
measured. The used antenna was a half-wave dipole encapsulated in an
iron sphere (made from 99.95{\%} iron, $\mu = 5000\mu _o $, $\sigma =
1.03\times 10^7$ S m$^{ - 1})$. The antenna impedance was $8.29\times
10^{ - 6} \ \Omega $ and the 9.9 mHz radiation emitted by the
antenna was completely absorbed by the iron along a critical thickness
0.110 m.

De Aquino mathematically predicted a dependence of the iron sphere 
gravitational mass on current, and experimentally proved that the total mass 
of the iron sphere (inertial mass of 60.5 kg) is nullified just at the 
current of 8.51 A. 

To rationalize this experiment case, instead of comparison of the
outputs expressed by equation (\ref{eq9}) rather energies are to be
related. The reason is a relatively long time of radiation passage
(about 16 s) through the iron layer.

The gravitational energy density associated with a 60.5 kg iron sphere with 
a 0.110 m radius is
\begin{equation}
\label{eq13}
\left| {\varepsilon _g } \right| = \frac{3m_{Fe} c^2}{4\pi  a   \delta 
^2}
\end{equation}
An amount of the gravitational energy in the sphere is 
\begin{equation}
\label{eq14}
\left| {E_g } \right| = \frac{m_{Fe} c^2\delta }{a}
\end{equation}
At interference of the gravitational and electromagnetic fields of iron 
sphere, each iron atom absorbs the energy $h\nu $ and thus the total 
electromagnetic energy in the iron sphere will be
\begin{equation}
\label{eq15}
E_{(h\nu )} = \frac{m_{Fe} }{m_{(atom)} }h\nu 
\end{equation}
where $m_{(atom)} $ is the mass of an iron atom. Identity of (\ref{eq14}) and (\ref{eq15}) 
leads to 
\begin{equation}
\label{eq16}
h\nu = \frac{m_{(atom)} c^2\delta }{a}
\end{equation}
Relation (\ref{eq16}) is valid when the gauge factor 
\begin{equation}
  \label{eq:38a}
 a = 1.399\times 10^{26} \mbox{ m} 
\end{equation}
which is in excellent agreement with its value calculated using the ENU 
model [12]. This value corresponds to the cosmological time 
\begin{equation}
  \label{eq:39a}
 t_c = 1.48\times 10^{10} \mbox{ years}  
\end{equation}

In this way, the de Aquino experiments and the ENU model provide an interval 
of the cosmological time 
\begin{equation}
  \label{eq:40a}
 t_c = (1.32 - 1.48)\times 10^{10} \mbox{ years} 
\end{equation}

In case of more experimental data available, the cosmological time can be 
calculated more precisely.

The above conclusions can be taken as a challenge both for those planning 
and performing experiments and those elaborating theoretical approaches. The 
given data manifest a mutual coherence of the Podkletnov's, de Aquino's and 
ours approaches. In addition, they bring a clear evidence of the necessity 
to exploit the general theory of relativity so as to understand and 
rationalize experiments devoted to gravitational mass changes.

\subsection*{References}
\begin{description}
\item {}
1. L. Smolin, Three Roads to Quantum Gravity, Basic Books, 2001

\item {}
2. E. Podkletnov, R. Nieminen, Physics C203 (1992) 441

\item {}
3. E. Podkletnov, xxx.lanl.gov/abs/cond-mat/9701074

\item {}
4. E. Podkletnov, G. Modanese, xxx.lanl.gov/abs/physics/0108005

\item {}
5. F. de Aquino, xxx.lanl.gov/abs/physics/0109060

\item {}
6. F. de Aquino, xxx.lanl.gov/abs/gr-qc/0005107

\item {}
7. F. de Aquino, xxx.lanl.gov/abs/gr-qc/0007069

\item {}
8. M. S\'{u}ken\'{\i}k, J. \v{S}ima, Spacetime {\&} Substance 2 (2001) 125

\item {}
9. S. Johnson, chemweb.com/physchem/0104003

\item {}
10. S. Johnson, chemweb.com/physchem/0103032

\item {}
11. V. Skalsk\'y, M. S\'{u}ken\'{\i}k, Astroph. Space Sci. 178 (1991) 169

\item {}
12. V. Skalsk\'y, M. S\'{u}ken\'{\i}k, Astroph. Space Sci. 215 (1994) 137

\item {}
13. V. Skalsk\'y, M. S\'{u}ken\'{\i}k, Astroph. Space Sci. 236 (1996) 295

\item {}
14. S. Davison, Corona, Univ. Oregon Press, 1999

\item {}
15. Compton's Encyclopeadia, Chicago, Compton's, 1996

\item {}
16. S.G. Lipson, H. Lipson, D.S. Tannhauser, Optical Physics (3$^{rd}$ ed.), 
Cambridge University Press, 1995

\item {}
17. F. de Aquino, J. New Energy, 5 (2000) 67

\item {}
18. M. S\'{u}ken\'{\i}k, J. \v{S}ima, Spacetime {\&} Substance 2 (2001) 79

\item {}
19. New Book of Knowledge, New York, Grolier, 1996, Vol. 17S

\item {}
20. F. de Aquino, xxx.lanl.gov/abs/physics/0205089
\end{description}

\end{document}